\documentclass[12pt]{article}

\ifx\pdfoutput\undefined
\usepackage[dvips,bookmarks]{hyperref}
\else
\usepackage{hyperref}
\fi
\hypersetup{colorlinks=false,bookmarksopen,bookmarksnumbered,citecolor=blue,
   pdfstartview=FitH}

\usepackage{latexsym}
\usepackage{amssymb,amsfonts,amsmath}
\usepackage{graphicx} 
\usepackage{indentfirst}
\usepackage{bbm}
\parskip=\medskipamount

\newcommand {\cD}{{\cal D}}
\newcommand {\cE}{{\cal E}}

\newcommand {\cK}{{\cal K}}
\newcommand {\cL}{{\cal L}}
\newcommand {\cM}{{\cal M}}
\newcommand {\cN}{{\cal N}}

\def\a{\alpha}
\def\b{\beta}
\def\c{\chi}
\def\d{\delta}
\def\e{\epsilon}
\def\f{\phi}
\def\g{\gamma}

\def\l{\lambda}
\def\m{\mu}
\def\n{\nu}
\def\o{\omega}

\def\q{\theta}

\def\s{\sigma}
\def\t{\tau}

\def\x{\xi}
\def\z{\zeta}

\def\F{\Phi}
\def\J{\Psi}
\def\L{\Lambda}
\def\O{\Omega}

\def\ri{{\rm i}}

\newcommand{\ad}{{\dot{\alpha}}}                           %new
\newcommand{\bd}{{\dot{\beta}}}                            %new
\newcommand{\ve}{\varepsilon}                            %new
\newcommand{\pa}{\partial}                           %new
\newcommand{\hf}{\frac12}
\newcommand{\be}{\begin{equation}}
\newcommand{\ee}{\end{equation}}
\newcommand{\bea}{\begin{eqnarray}}
\newcommand{\eea}{\end{eqnarray}}
\newcommand{\non}{\nonumber}
\newcommand{\ba}{\begin{array}}
\newcommand{\ea}{\end{array}}

\newcommand{\bm}[1]{\mbox{\boldmath$#1$}}

\def\double #1{#1{\hbox{\kern-2pt $#1$}}}

\newcommand{\gd}{{\dot\g}}

\newcommand{\bsubeq}{\begin{subequations}}
\newcommand{\esubeq}{\end{subequations}}

\newcommand{\eps}{{\epsilon}}

\newcommand{\dalpha}{{\dot{\alpha}}}

\newcommand{\btheta}{{\bar\theta}}
\newcommand{\N}{{\mathcal N}}
\newcommand{\eol}{\notag \\}
\newcommand{\rd}{\mathrm d}
\newcommand{\veps}{\varepsilon}

\begin{document}

\begin{titlepage}
\begin{flushright}
May 2011 \\ 
Revised version: August 2011\\
\end{flushright}
\vspace{5mm}

\begin{center}
{\Large \bf $\bm{\cN = 2}$  supersymmetric sigma-models in AdS}
\\ 
\end{center}

\begin{center}

{\bf Daniel Butter and  Sergei M. Kuzenko }

\footnotesize{
{\it School of Physics M013, The University of Western Australia\\
35 Stirling Highway, Crawley W.A. 6009, Australia}}  ~\\
\texttt{dbutter,\,kuzenko@cyllene.uwa.edu.au}\\
\vspace{2mm}

\end{center}
\vspace{5mm}

\begin{abstract}
\baselineskip=14pt
We construct  the most general $\cN=2$ supersymmetric nonlinear sigma-model in 
four-dimensional anti-de Sitter (AdS) space in terms of $\cN=1$ chiral superfields. 
The target space is shown to be a non-compact hyperk\"ahler manifold
restricted to possess a special 
Killing vector field. A remarkable property of the sigma-model constructed is that 
the algebra of ${\rm OSp(2|4)}$ transformations is closed off the mass shell.
\end{abstract}
\vspace{0.5cm}

\vfill
\end{titlepage}

\newpage
\renewcommand{\thefootnote}{\arabic{footnote}}
\setcounter{footnote}{0}

%\tableofcontents

%\pagestyle{plain}
%\pagenumbering{arabic}

% Equation numbers
\numberwithin{equation}{section}

%\allowdisplaybreaks

%===BEGIN DOCUMENT=============================================================
%\newpage

\section{Introduction}

In 1986, Hull et al. \cite{HKLR} formulated, building on the earlier work of
Lindstr\"om and Ro\v{c}ek  \cite{LR},  
general four-dimensional $\cN=2$ rigid supersymmetric sigma-models 
(without superpotentials) in terms of $\cN=1$ chiral superfields. 
In 2006, the approach of \cite{HKLR} was extended to include superpotentials
\cite{BX}.\footnote{The reference \cite{BX} also considered the lift of these
models to 5D $\N=1$ supersymmetry. The case of 6D $\cN=(1,0)$ supersymmetry
was further studied in \cite{Gates:2006it}.}
The most general $\cN=2$ superconformal sigma-models have been formulated in terms 
of $\cN=1$ chiral superfields  only recently in \cite{K-duality}.\footnote{The main virtue of 
the $\cN=1$ superspace formulations \cite{HKLR,BX,K-duality} is that one of the two supersymmetries 
is realized off-shell. The analogous component results appeared earlier.
Specifically,  the rigid supersymmetric 
sigma-models with eight supercharges were first constructed in \cite{A-GF}. The construction
of \cite{A-GF} was extended to include
a superpotential in \cite{GTT}. General $\cN=2$ rigid superconformal sigma-models were 
studied in \cite{deWKV0,deWKV}.}
The formulation given in \cite{HKLR} is rather geometric, for it makes use of the geometric structures
that are intrinsic to the hyperk\"ahler target space. 

In this paper, our aim is to construct the most general $\cN=2$ AdS supersymmetric 
sigma-models\footnote{General  off-shell $\cN=2$ AdS supersymmetric 
sigma-models have already been formulated in the $\cN=2$ AdS superspace in \cite{KT-M-ads},  
building on the projective-superspace formulation for $\cN=2$ supergravity-matter 
systems \cite{KLRT-M1,KLRT-M2}. Using the off-shell $\cN=2$ sigma-model actions of \cite{KT-M-ads}, 
one can in principle derive their reformulation in terms of $\cN=1$ chiral superfields upon 
(i) eliminating the (infinitely many) auxiliary superfields; and (ii) performing superspace duality transformations.
However, these two technical procedures are quite difficult to implement explicitly in general.}
in terms of covariantly chiral superfields on $\cN=1$ AdS superspace.\footnote{Historically, the $\cN=1$ AdS superspace,  AdS$^{4|4}:={\rm OSp(1|4) }/  {\rm O(3,1)}$,
was introduced in \cite{Keck,Zumino}, and the superfield approach to ${\rm OSp(1|4)}$ supersymmetry
was developed by Ivanov and Sorin \cite{IS}.}
Achieving this goal proves to require a more involved analysis than that given in 
the rigid supersymmetric case \cite{HKLR,BX,K-duality}, simply because the superspace geometry is curved 
(even if maximally symmetric).  We carry out such an analysis, and its outcome 
turns out to be  really rewarding.
We prove that the $\cN=2$ AdS supersymmetric sigma-models constructed are {\it off-shell}, 
that is the algebra of the ${\rm OSp(2|4)}$ transformations closes off the mass shell.
Moreover, the target space is shown to be a non-compact hyperk\"ahler manifold restricted
to possess a special Killing vector field which rotates the complex structures.

This paper is organized as follows. In section 2, we briefly review the properties of AdS
nonlinear sigma-models in $\N=1$ superspace. Then in section 3 we present the conditions
that the $\N=1$ action must obey in order to possess a second supersymmetry. In section 4,
we give the component formulation of this action and the action of $\N=2$
supersymmetry on its component fields. In section 5, we elaborate upon several
interesting implications of our results.

\section{$\cN=1$ nonlinear sigma-models in AdS}

Before discussing supersymmetric nonlinear sigma-models, 
it is worth giving  essential  information about the $\cN=1$ superspace AdS$^{4|4}$
(see \cite{BK} for more details) which is a maximally symmetric solution of old minimal 
supergravity with a cosmological term. The corresponding covariant derivatives,\footnote{We
follow the notation and conventions adopted in \cite{BK}, except we use lower
case Roman letters for tangent-space vector indices.} 
\bea
\cD_{\rm A}=(\cD_{\rm a}, \cD_\a,{\bar \cD}^\ad)=E_{\rm A}{}^{M}\pa_M+\hf\f_{\rm A}{}^{\rm bc}M_{\rm bc}~,
\eea
obey the following (anti-)commutation relations:
\begin{subequations}
\bea
&&\{\cD_\a,\cD_\b\}=
-4\bar{\mu}M_{\a\b}~, \qquad ~~~
\{\cD_\a,{\bar \cD}_\bd\}=-2\ri(\s^{\rm c})_{\a \bd}\cD_{\rm c} \equiv -2\ri \cD_{\a\bd}
~,~~
\label{N=1-AdS-algebra-1}
\\
&&
{[}\cD_{\rm a},\cD_\b{]}=
-\frac{\ri}{ 2}   {\bar \mu} ({\s}_{\rm a})_{\b\gd} {\bar \cD}^\gd~,\qquad
{[}\cD_{\rm a},\cD_{\rm b}{]}=-| { \mu}|^2 M_{\rm ab}~,~~~~~~
\label{N=1-AdS-algebra-2}
\eea
\end{subequations}
with $\m$ a complex non-vanishing parameter which can be viewed as
a square root of the curvature of the anti-de Sitter space. 
The ${\rm OSp(1|4)}$ isometries of AdS$^{4|4}$ are generated by Killing vector fields
defined as 
\bea
\L=\l^{\rm a} \cD_{\rm a}+\l^\a \cD_\a+{\bar \l}_\ad {\bar \cD}^\ad~, \qquad
{[}\L+\hf \o^{\rm bc}M_{\rm bc},\cD_{\rm A}{]}=0~,
\label{N=1-killings-0}
\eea
for some Lorentz transformation generated by   $\o^{\rm bc}$. 
As shown in \cite{BK}, 
the equations in (\ref{N=1-killings-0}) are equivalent to 
\begin{subequations}\label{2.4}
\bea
\cD_{(\a}\l_{\b)\bd}&=&0~, \qquad  {\bar \cD}^\bd\l_{\a\bd}   + 8\ri\l_\a=0~,\\
\cD_\a\l^\a&=&0~,
\qquad
{\bar \cD}_\ad\l_\a  + {\ri\over 2}{\mu}\l_{\a\ad}  =0~, \\
\o_{\a\b}&=&\cD_\a\l_\b~.
\eea
\end{subequations}

The most general  nonlinear sigma-model in $\N=1$ AdS superspace is given by
\begin{align}
S = \int \rd^4x\, \rd^4\theta \, E\, \cK (\f^a, \bar \f^{\bar b})~,
\label{sigma-action}
\end{align}
where $E^{-1}= {\rm Ber}\, (E_{\rm A}{}^M) $.
The dynamical variables $\f^a$ are covariantly chiral superfields, ${\bar \cD}_\ad \f^a =0$,
and at the same time local complex coordinates of a complex  manifold $\cM$.
Unlike in the Minkowski case, the action does not possess K\"ahler
invariance since 
\begin{align}\label{2.6}
\int \rd^4x\, \rd^4\theta \, E\, F (\f^a)= \int \rd^4x\, \rd^2\theta \, \cE\, \mu F (\f^a)  \neq 0~,
\end{align}
with $\cE$ the chiral density. Nevertheless, 
K\"ahler invariance naturally emerges if we represent the Lagrangian as
\bea\label{eq_KW}
\cK (\f , \bar \f ) = K(\f , \bar \f) + \frac{1}{\m} W(\f) +  \frac{1}{\bar \m} \bar W( \bar \f) ~, 
\eea
for some K\"ahler potential $K$ and superpotential $W$. 
Under a K\"ahler transformation, these transform as
\begin{align}\label{eq_AdSKahler}
K \rightarrow K + F + \bar F, \qquad
W \rightarrow W - \mu F~.
\end{align}
The K\"ahler metric defined by 
\begin{align} \label{Kahler_metric}
g_{a \bar b} := \partial_a \partial_{\bar b} \cK = \partial_a \partial_{\bar b} K
\end{align}
is obviously invariant under (\ref{eq_AdSKahler}).

The nonlinear sigma-model (\ref{sigma-action}) is manifestly invariant under 
arbitrary $\cN=1$ AdS isometry transformations 
\bea
\d_\L \f^a = \L \f^a~, 
\eea
with the operator $\L$ defined by eqs. (\ref{N=1-killings-0}) and (\ref{2.4}).

Because of (\ref{2.6}), the Lagrangian $\cK$ in  (\ref{sigma-action}) should be  a globally defined function 
on the K\"ahler target space $\cM$. This implies that the K\"ahler two-form, 
 $ \O=\ri \,g_{a \bar b} \, \rd \f^a \wedge \rd \bar \f^{\bar b}$,  associated with 
(\ref{Kahler_metric}), is exact and hence  $\cM$ is necessarily non-compact. 
We see that the sigma-model couplings in AdS are more restrictive than in the Minkowski case.
The same conclusion follows from our recent analysis of AdS supercurrent multiplets \cite{BK2011}.
In \cite{BK2011} we demonstrated that $\cN=1$ AdS supersymmetry allows the existence 
of just one minimal ($12+12)$ supercurrent, unlike the case of Poincar\'e supersymmetry 
admitting three ($12+12)$ supercurrents. The corresponding AdS supercurrent 
is associated with the old minimal supergravity   and coincides with  the AdS extension of the Ferrara-Zumino 
multiplet \cite{FZ}. An immediate application of this result is that all supersymmetric sigma-models in AdS
must possess a well-defined  Ferrara-Zumino  multiplet. The same conclusion also follows
from the exactness of $\O$ and earlier results of Komargodski and Seiberg \cite{KS}
who demonstrated that all {\it rigid} supersymmetric sigma-models with an exact K\"ahler two-form 
possess a well-defined  Ferrara-Zumino  multiplet. 
The exactness of $\O$ for the general $\cN=1$ sigma-models in AdS 
has independently been observed in recent publications
\cite{Adams:2011vw} and \cite{Festuccia:2011ws} which appeared shortly after \cite{BK2011}.

We should discuss briefly how the structure \eqref{sigma-action}
emerges within a supergravity description. 
(Our discussion here is similar to that recently given in \cite{Adams:2011vw}.) 
Recall that nonlinear sigma-models
may be coupled to supergravity via
\begin{align}
S = -\frac{3}{\kappa^2} \int \rd^4x\, \rd^4\theta \, E\, e^{-\kappa^2 K / 3}
     + \int \rd^4x\, \rd^2\theta \,  \cE\, W_{\rm sugra}
     + \int \rd^4x\, \rd^2\btheta \, \bar\cE\, \bar W_{\rm sugra}
\end{align}
where the K\"ahler potential $K$ and the superpotential $W_{\rm sugra}$
transform under K\"ahler transformations as
\begin{align}\label{eq_SGKahler}
K \rightarrow K + F + \bar F, \quad
W_{\rm sugra} \rightarrow e^{-\kappa^2 F} W_{\rm sugra}~.
\end{align}
The parameter $\kappa$ corresponds to the inverse Planck mass.
To derive an AdS model from a supergravity model, we insert a
cosmological term by hand in the superpotential
\begin{align}
W_{\rm sugra} = \frac{\mu}{\kappa^2} + W
\end{align}
and consider the limit of small $\kappa$. The terms which diverge in
such a limit correspond to pure supergravity with a cosmological
constant and the supergravity equations of motion may be solved
to yield an AdS solution, freezing the supergravity structure.
The terms which remain as $\kappa$ tends to zero can be shown to
take the form \eqref{sigma-action} with \eqref{eq_KW}. The
corresponding limit of \eqref{eq_SGKahler} yields \eqref{eq_AdSKahler}.

\section{$\cN=2$  nonlinear sigma-models in AdS}
We now  turn to implementing our main goal, that  is to look for those restrictions on the target space geometry 
which guarantee that  the theory (\ref{sigma-action}) is  $\cN=2$ supersymmetric.

\subsection{$\N=2$ supersymmetry transformations}
We make the following ansatz\footnote{The transformation law (\ref{N=2SUSYtr}) is a generalization 
of that derived in \cite{KT-M-ads}, using manifestly $\cN=2$ supersymmetric techniques, 
in the case of  a free $\cN=2$ hypermultiplet $\f^a = (\F, \J)$ for which 
$\d_\ve \F =  \frac{1}{2} (\bar\cD^2 - 4 \mu) (\veps \bar \J)$ and 
$\d_\ve \J =  - \frac{1}{2} (\bar\cD^2 - 4 \mu) (\veps \bar \F)$.
The ansatz (\ref{N=2SUSYtr}) also has a correct  super-Poincar\'e limit \cite{HKLR,BX}. } 
for the action of a second
supersymmetry on the chiral superfield $\phi^a$:
\begin{align}
\delta_\ve \phi^a = \frac{1}{2} (\bar\cD^2 - 4 \mu) (\veps \bar \Omega^a)
\label{N=2SUSYtr}
\end{align}
where $\bar\Omega^a$ is a function of $\phi$ and $\bar\phi$ which has to be determined.
The parameter $\veps$ is real, $\bar \ve = \ve$,  and  constrained to obey \cite{GKS}
\begin{align}
(\bar \cD^2 - 4 \mu) \veps = \bar\cD_\dalpha \cD_\alpha \veps = 0\quad \Longrightarrow \quad
\cD_{\a\ad} \ve = 0~.
\label{epsilon-constraints}
\end{align}
Defining $\ve_\a :=\cD_\a \ve$, the second constraint implies that $\ve_\a$ is chiral, ${\bar \cD}_\ad \ve_\a=0$.
The parameter $\ve$ naturally originates within the $\cN=2$ AdS superspace approach \cite{KT-M-ads}.
The isometries of $\cN=2$ AdS superspace are described by the corresponding Killing vector fields
defined in \cite{KT-M-ads}.  Upon reduction to $\cN=1$ AdS superspace, any $\cN=2$ Killing vector 
 produces an $\cN=1$ Killing vector  $\L$, eq. (\ref{N=1-killings-0}),  and $\ve$.

The $\q$-dependent parameter $\ve$, due to the constraints eq. (\ref{epsilon-constraints}), 
contains two components:
(i)  a  bosonic parameter $\x$ which is defined by $ \ve |_{\q=0} = \x  |\m|^{-1} $ 
and describes  the O(2) rotations;  
and (ii) a fermionic parameter $\e_\a := \cD_\a \ve|_{\q=0} $ along with its conjugate, 
which generate the second supersymmetry. Schematically, the $\ve$ looks like
\bea
\ve \sim \frac{\x}{|\m|} + \e^\a \q_\a + \bar \e_\ad \bar \q^\ad 
-\x \Big( \frac{\bar \m}{|\m |} \q^2 +  \frac{ \m}{|\m |} \bar \q^2 \Big) ~.
\label{SUSY_parameter}
\eea

On the mass shell, the right-hand side of (\ref{N=2SUSYtr})
should transform as a vector field of type (1,0) under
reparametrizations of the target space.
Due to the constraints (\ref{epsilon-constraints}), the transformation $\delta \phi^a$ may be
rewritten in the form 
\begin{align}
\delta \phi^a = \bar\veps_\dalpha \bar\cD^\dalpha \bar\Omega^a
     + \frac{1}{2} \veps \,\bar\cD^2 \bar\Omega^a
     \label{N=2SUSYtr2}
\end{align}
which makes clear that $\bar\Omega^a$ is defined only up to a holomorphic vector,
\begin{align}
\bar\Omega^a \rightarrow \bar\Omega^a + H^a(\phi)~.
\label{3.5}
\end{align}

\subsection{Conditions for $\cN=2$ supersymmetry}
We turn to discussing the conditions for the sigma-model  action (\ref{sigma-action}) 
to be invariant under the $\cN=2$ 
supersymmetry transformations (\ref{N=2SUSYtr}) and (\ref{epsilon-constraints}).

A large amount  of information can be extracted from the following requirement
\bea
\frac{\d}{\d \f^a} \int \rd^4x\, \rd^4\theta\, E\, 
\Big\{ \cK_{b} (\bar \cD^2 - 4 \mu) (\veps \bar\Omega^b)
     +  \cK_{\bar b} (\cD^2 - 4 \bar \mu) (\veps \Omega^{\bar b})
     \Big\} =0
     \label{3.7}
\eea
which must hold if the action is invariant. This requirement is technically easier to analyse 
than the invariance condition $\d S=0$. The technical details of such an analysis will be reported elsewhere.
Here we only present the final results.
As in the globally supersymmetric
case \cite{HKLR}, one may introduce a tensor $\omega_{\bar a \bar b}$ via
\begin{align}
\omega_{\bar a \bar b} := g_{\bar a c} \,\bar\Omega^c{}_{,\bar b}~,
     \qquad
\bar\Omega^c{}_{,\bar b} := \partial_{\bar b} \bar\Omega^c~.
\label{3.8}
\end{align}
Eq. (\ref{3.7}) implies that $\omega_{\bar a \bar b}$ is both a two-form,
\bea
\omega_{\bar a \bar b}      = -\omega_{\bar b \bar a}~,
\label{3.9}
\eea
 and covariantly constant,
\begin{align}
\nabla_c \,\omega_{\bar a \bar b} = 0, \qquad
\nabla_{\bar c} \,\omega_{\bar a \bar b} = 0~,
\label{3.10}
\end{align}
and similarly for its complex conjugate $\omega_{ab}$.\footnote{It was shown in 
\cite{Kuzenko:2010rp} that if the tensor $\omega_{\bar a \bar b}$ defined by (\ref{3.8}) is 
antisymmetric, then the second equation in  (\ref{3.10}) automatically holds.}
These conditions imply that both
$\omega_{a b}$ and $\omega^{ab} := g^{a \bar a} g^{b \bar b} \omega_{\bar a \bar b}$
are holomorphic, $\o_{ab} = \o_{ab}(\f)$ and  $\o^{ab} = \o^{ab}(\f)$.

The conditions (\ref{3.8}) and (\ref{3.9}) are exactly the same as in the
rigid supersymmetric case \cite{HKLR}. There is in addition one extra
purely AdS condition that follows from (\ref{3.7}). We find that the following
equation must hold:
\begin{align}
 \mu \,\partial_a \left(g_{\bar a c} \omega^{cb} \cK_b\right)
     + \bar\mu \,\partial_{\bar a} \left(g_{a \bar c} \omega^{\bar c\bar b} \cK_{\bar b}\right) =0~.
\end{align}
If we define the vector field 
\begin{align}
V^\mu = (V^a, V^{\bar a})~,\qquad
V^a := \hf \frac{\mu}{|\m|} \omega^{ab} \cK_b, \qquad
V^{\bar a} := \hf \frac{\bar\mu}{|\m |} \omega^{\bar a \bar b} \cK_{\bar b}~,
\label{vector_field}
\end{align}
then the above equation may be written as
\begin{align}
 \nabla_a V_{\bar b} + \nabla_{\bar b} V_a =0~.
\label{3.13}
\end{align}
In addition, since $\nabla_a V_b = - \bar\mu \omega_{ab} /2 |\mu|$, we also have 
\begin{align}
 \nabla_a V_{b} + \nabla_{b} V_a =0~.
 \label{3.14}
\end{align}
Together, these conditions imply that $V= V^a \pa_a +\bar V^{\bar a} \pa_{\bar a}$ 
is a Killing vector field on the K\"ahler  manifold. 
By construction it also obeys
\begin{align}
V^a \partial_a \cK = V^{\bar a} \partial_{\bar a} \cK = 0~.
\end{align}

This Killing vector turns out to also obey one additional critical property:
it acts as a rotation on the three complex structures! One can easily show that
the Lie derivative of the complex structures are given by
\begin{subequations}\label{eq_CSrotate}
\begin{align}
\cL_V J_1 &= \frac{\textrm{Im}\,\mu}{|\mu|} \, J_3 = \sin\theta J_3 \\
\cL_V J_2  &= -\frac{\textrm{Re}\,\mu}{|\mu|} J_3 = -\cos\theta J_3 \\
\cL_V J_3  &= \frac{\textrm{Re}\,\mu}{|\mu|}J_2 - \frac{\textrm{Im}\,\mu}{|\mu|} J_1
	= \cos\theta J_2 - \sin\theta J_1
\end{align}
\end{subequations}
where $\theta = \arg\mu$. In particular, the specific linear combination
$J_1 \cos\theta + J_2 \sin\theta$ turns out to be invariant under the
Lie derivative. This condition is remarkable since it implies that $V^\mu$ is
holomorphic with respect to this specific complex structure,\footnote{In other words,
if one were to work in coordinates where $J_1 \cos\theta + J_2 \sin\theta$
is diagonalized to $\textrm{diag}(\ri \mathbbm 1_n, -\ri \mathbbm 1_n)$, then $V^\mu$ would be
holomorphic in the usual sense.} but not \emph{tri-holomorphic}. These features
have recently been observed in two papers \cite{BX2011,BaggerLi}
in which supersymmetric nonlinear sigma-models in ${\rm AdS}_5$ were formulated
first in terms of 4D $\cN=1$ chiral superfields \cite{BX2011} and then involving
component fields  \cite{BaggerLi}. As argued in \cite{BX2011}, 
the  ${\rm AdS}_5$ supersymmetry requires the sigma-model target space to be hyperk\"ahler and
possess a holomorphic Killing vector field.\footnote{The Killing vector turns out to
be holomorphic due to a certain embedding of the hypermultiplets into 4D $\N=1$
chiral superfields.} In that case, the Killing
vector field is again holomorphic with respect to just one of the complex
structures, but not tri-holomorphic.

The above properties follow solely from the requirement (\ref{3.7}), 
without a direct analysis of the invariance condition $\d S=0$. 
However, taking into account the properties (\ref{3.9}), (\ref{3.10})
and (\ref{3.13}), (\ref{3.14}) it can be shown that the action is indeed invariant.  
We shall describe the derivation in a separate publication.

As a simple example, consider the $\cN=2$ linear sigma-model \cite{KT-M-ads}
\bea
S= \int \rd^4 x \, {\rm d}^4\q \, E\Bigg(
\bar{{ \F}}{ \F}
+\bar{{ \Psi}}{ \Psi}
-\ri\frac{\bar \mu}{|\mu |} \Big(1+\frac{m}{|\m|}\Big){ \Psi}{ \F} 
+\ri\frac{\mu}{|\mu |} \Big(1+\frac{m}{|\m|}\Big) {\bar { \Psi}} {\bar { \F}}
\Bigg) ~,~~~
\label{S-CC-massive}
\eea
with $\f^a=(\F, \J)$ covariantly chiral superfields, and $m$ a mass parameter  
(the choice $m=- |\m|$ corresponds to the superconformal massless hypermultiplet).
Using the explicit expression for the holomorphic two-form 
\begin{align}
\o_{ab} = \left(\begin{array}{cc}
0 & 1 \\
-1 & 0
\end{array}\right)~,
\end{align}
it is easy to check that the vector field (\ref{vector_field}) given
by $V_a = (V_\phi, V_{\psi})$ and $V_{\bar a} = (V_{\bar\phi}, V_{\bar\psi})$  with
\begin{subequations}
\begin{align}
V_{\phi} = \frac{1}{2} \frac{\bar\mu}{|\mu|} \Psi + \frac{\ri}{2}  \left(1 + \frac{m}{|\mu|}\right) \bar\Phi~, \qquad
V_{\psi} = -\frac{1}{2} \frac{\bar\mu}{|\mu|} \Phi - \frac{\ri}{2}  \left(1 + \frac{m}{|\mu|}\right) \bar\Psi  \\
V_{\bar \phi} = \frac{1}{2} \frac{\mu}{|\mu|} \bar\Psi - \frac{\ri}{2}  \left(1 + \frac{m}{|\mu|}\right) \Phi~, \qquad
V_{\bar \psi} = -\frac{1}{2} \frac{\mu}{|\mu|} \bar\Phi + \frac{\ri}{2}  \left(1 + \frac{m}{|\mu|}\right) \Psi 
\end{align}
\end{subequations}
indeed obeys (\ref{3.13}) and (\ref{3.14}).

It should be remarked that, modulo transformations (\ref{3.5}),  we can choose
\bea
\bar \O^a (\f , \bar \f ) = \o^{ a b} (\f) \, \cK_b (\f , \bar \f)~,
\eea
similarly to the super-Poincar\'e case \cite{HKLR}. 
The specific feature of the AdS case
is that $ \cK_b $ is a one-form, and thus $\bar \O^a $ is necessarily a vector field.
Comparing the expression for $\bar \O^a$ with (\ref{vector_field}) shows  that $\bar \O^a \propto V^a$.

\subsection{Closure of the supersymmetry algebra}
Let us calculate the commutator of two second supersymmetry transformations (\ref{N=2SUSYtr}).
This calculation is rather short and the result is 
\bea
[ \delta_{\ve_2} , \delta_{\ve_1}  ]\phi^a = - \omega^{ac} \omega_{cb} 
\Big( -\hf \tilde{\l}^{\a\ad} \cD_{\a \ad} + \tilde{\l}^\a \cD_\a\Big) \f^b ~,
\eea
where 
\bea
\tilde{\l}^{\a\ad} &:=& 4\ri \big(\ve^\a_2 \bar \ve^\ad_1 - \ve^\a_1 \bar \ve^\ad_2 \big)~, \qquad 
\tilde{\l}^\a:= 2\m \big( \ve^\a_1 \ve_2 - \ve^\a_2 \ve_1\big)
\eea
are the components of the first-order operator
$\L_{[\ve_2, \ve_1]} = -\hf \tilde{\l}^{\a\ad} \cD_{\a\ad} + \tilde{\l}^\a \cD_\a + \bar{\tilde{\l}}_\ad \bar \cD^\ad $
which proves to be an AdS Killing vector field, see eqs. (\ref{N=1-killings-0}) and (\ref{2.4}).
If we impose
\begin{align}
\omega^{ac} \omega_{cb} = -\delta^a{}_b~, 
\label{3.18}
\end{align}
then the above result turns into
\bea
[ \delta_{\ve_2} , \delta_{\ve_1}  ]\phi^a = \L_{[\ve_2, \ve_1]}  \f^a ~.
\label{3.19}
\eea

We see from (\ref{3.19}) that the commutator $[ \delta_{\ve_2} , \delta_{\ve_1}  ]\phi^a $
closes off the mass shell. 
This is similar to the supersymmetry structure  
within the Bagger-Xiong formulation \cite{BX} for $\cN=2$ rigid supersymmetric sigma-models.
However, in the case of flat superspace, the commutator of the first and the second supersymmetries 
closes {\it only} on-shell \cite{BX}.  What about the AdS case? Computing 
the commutator of the $\cN=1$ AdS transformation and the second supersymmetry 
transformation gives
\bea
[\d_\L , \d_\ve ] \f^a 
= - \frac{1}{2} (\bar\cD^2 - 4 \mu) 
\Big ((\L  \veps )\bar \Omega^a \Big )~.
\eea 
Since $\L$ is an $\cN=1$ Killing vector field, the parameter $\ve'= \L \ve$ obeys the constraints 
(\ref{epsilon-constraints}) and hence generates a second supersymmetry transformation. 
We observe that commuting 
the $\cN=1$ AdS transformation and the second supersymmetry gives
a second supersymmetry transformation, 
\bea
[\d_\L , \d_\ve ] \f^a 
= - \d_{\L\ve} \f^a ~.
\eea
As a result, the algebra of ${\rm OSp(2|4)}$ transformations is closed 
off the mass shell!\footnote{It should be mentioned that the linearized action for {\it all} massless supermultiplets
of arbitrary superspin in $\cN=1$ AdS superspace \cite{GKS} is also invariant under 
$\cN=2$  supersymmetry transformations which close off-shell. }

Let us return to the equation (\ref{3.18}). Its implications are the same 
as in the super-Poincar\'e case \cite{HKLR}.
In addition to the canonical complex structure
\begin{align}\label{complex_structure1} 
J_3 = \left(\begin{array}{cc}
\ri \,\delta^a{}_b & 0 \\
0 & -\ri \,\delta^{\bar a}{}_{\bar b}
\end{array}\right),
\end{align}
we may construct two more using $\omega^a{}_{\bar b}$
\begin{align}
\label{complex_structure2}
J_1 = \left(\begin{array}{cc}
0 & \omega^a{}_{\bar b} \\
\omega^{\bar a}{}_b & 0
\end{array}\right), \qquad
J_2 = \left(\begin{array}{cc}
0 & \ri\, \omega^a{}_{\bar b} \\
-\ri\, \omega^{\bar a}{}_b & 0
\end{array}\right)
\end{align}
such that $\cM$ is K\"ahler with respect to each of them.
The operators $J_A = (J_1, J_2, J_3)$ obey the quaternionic algebra
\begin{align}
J_A J_B = -\delta_{A B} \mathbb{I} + \eps_{ABC} J_C~.
\end{align}
Thus, $\cM$ is a hyperk\"ahler manifold. 
In accordance with the discussion in section 2, this manifold is non-compact. 
The above analysis also shows that $\cM$ must possess a special Killing vector.

Using (\ref{3.18}), it is easy to establish the equivalence 
\bea
(\bar \cD^2 - 4 \mu) \cK_a =0 \quad \Longleftrightarrow \quad (\bar \cD^2 - 4 \mu) (\o^{ab}\cK_b)=0~.
\eea
This results implies that the following rigid symmetry of the $\cN=2$ sigma-model 
\bea
\d \f^a = \z (\bar \cD^2 - 4 \mu) (\o^{ab}\cK_b) ~, \qquad \z \in {\mathbb C} 
\eea
is trivial.

It is well-known that when $\N=2$ sigma-models are coupled to supergravity,
their target spaces must be quaternionic K\"ahler manifolds \cite{BW}.
Unlike the hyperk\"ahler spaces which are Ricci-flat, their quaternionic K\"ahler 
cousins are Einstein spaces with a non-zero constant scalar curvature
(see, e.g., \cite{Besse} for a review).
Since AdS is a curved geometry, one may wonder whether the target spaces
of $\cN=2$ sigma-models in AdS should also be   quaternionic K\"ahler.
Yet we have shown here that within AdS, the geometry is hyperk\"ahler
just as in Minkowski space. The reason is simple. As shown in
\cite{BW}, the  scalar curvature in the target space of locally supersymmetric sigma-models 
must be nonzero and proportional to $\kappa^2$,
\begin{align}
R = -8 \kappa^2 (n^2 + 2n)~,
\end{align}
where the real dimension of the target space is $4n$.
But AdS (or Minkowski) space can be interpreted
as the $\kappa^2 \rightarrow 0$ limit of supergravity with (or without) a cosmological
constant $\mu$. In such a limit, we find indeed that the quaternionic
K\"ahler requirement from supergravity reduces to a hyperk\"ahler requirement.

\subsection{$\cN=2$ superconformal sigma-models}
Both Minkowski and AdS $\cN=2$ superspaces have the same superconformal group 
$\rm SU(2,2|2)$. Thus all $\cN=2$ rigid superconformal sigma-models should be invariant under
the $\cN=2$ AdS supergroup $\rm OSp(2|4)$. Here we elaborate on this point.

Target spaces for $\cN=2$ superconformal sigma-models are hyperk\"ahler cones
(see \cite{deWRV} and references therein). A hyperk\"ahler cone is a hyperk\"ahler manifold
possessing a homothetic conformal Killing vector field. 
Let us recall the salient facts about  
homothetic conformal Killing vector fields (see \cite{deWRV,GR} for more details).
By definition, a homothetic conformal Killing vector field $\c$
on a K\"ahler manifold $(\cM, g_{a\bar b})$, 
\bea
\c = \c^a \frac{\pa}{\pa \f^a} + {\bar \c}^{\bar a}  \frac{\pa}{\pa {\bar \f}^{\bar a}}
\equiv \c^\m \frac{\pa}{\pa \f^\m} ~,
\eea
obeys the constraint
\bea
\nabla_\n \c^\m = \d_\n{}^\m \quad \Longleftrightarrow \quad 
\nabla_b \c^a = \d_b{}^a~, \qquad 
\nabla_{\bar b} \c^a = \pa_{\bar b} \c^a = 0~.
\label{hcKv}
\eea
In particular,  $\c $ is holomorphic. Its properties include:
\bea
{ g}_{a \bar b} \, \c^a {\bar \c}^{\bar b} ={ \cK}~, 
\qquad \c_a := {g}_{a \bar b} \,{\bar \c}^{\bar b} = \pa_a {\cK}~,
\label{hcKv-pot}
\eea
with $\cK$ the K\"ahler potential. If $\cN=2$ superconformal sigma-models are realized in 
$\cN=1$ Minkowski superspace, the second supersymmetry is given in terms of $\c$ \cite{K-duality}.

We have to show that the above properties of $\c$
imply the existence of a Killing vector field
\bea
V^\m = (V^a, V^{\bar a})=
\frac{1}{2|\m|} \Big( \mu\, \omega^{ab} \cK_b\, , \,\bar\mu \, \omega^{\bar a \bar b} \cK_{\bar b}\Big)
=  \frac{1}{2|\m|} \Big( \mu\, \omega^{ab} \c_b\, , \,\bar\mu \, \omega^{\bar a \bar b} \c_{\bar b}\Big)~,
\label{3.31}
\eea
for any non-zero complex parameter $\m$. By representing $ 2|\m | V_a = \bar\m\,  \o_{ab}\c^b$ and 
using the facts that $\o_{ab}$ and $\c^b$ are holomorphic, the condition (\ref{3.13}) follows. 
The other condition, eq.  (\ref{3.14}), holds automatically. 

It is instructive to give a slightly different proof that (\ref{3.31}) is a Killing vector, which shows that $V^\mu$  
belongs to the Lie algebra of the group SU(2) isometrically acting on the hyperk\"ahler cone.
As shown e.g. in \cite{deWRV,GR}, associated with the complex structures $(J_A)^\m{}_\n$, eqs. 
(\ref{complex_structure1}) and (\ref{complex_structure2}),  are   
the three Killing vectors $X_A^\m := (J_A)^\m{}_\n \c^\n$ which span  the Lie algebra of SU(2).
In particular, we have that $(J_1)^\m{}_\n \c^\n = (\o^{ab}\cK_b\, , \, \omega^{\bar a \bar b} \cK_{\bar b})$
and $(J_2)^\m{}_\n \c^\n = (\ri \, \o^{ab}\cK_b\, , -\ri\,  \omega^{\bar a \bar b} \cK_{\bar b})$
are Killing vectors. The Killing vector (\ref{3.31}) is simply a real combination of  $(J_1)^\m{}_\n \c^\n$
and  $(J_2)^\m{}_\n \c^\n$, and thus $V^\m$ belongs   the Lie algebra of SU(2).

\newcommand{\ccD}{{\bm \cD}}
\section{$\N=2$ nonlinear sigma-models in components}

We turn now to the component description of $\N=2$ non-linear sigma-models
in AdS. The evaluation of the superspace action is straightforward, and
makes use of the $\cN=1$ AdS reduction rule (see e.g. \cite{KS94} or standard texts on $\cN=1$ 
supergravity \cite{BK,WB})
equivalent to 
\begin{align}
S &= \int \rd^4x\, \rd^4\theta\, E\, \cK \non\\
     &= \int \rd^4x\, e\, \Bigg\{
     \frac{1}{16} \cD^\alpha (\bar\cD^2-4\mu) \cD_\alpha \cK
     - \frac{1}{4} \bar\mu \bar\cD^2 \cK
     - \frac{1}{4} \mu \cD^2 \cK
     + 3 \mu \bar \mu \cK
     \Bigg\}
\end{align}
where $E^{-1} = {\rm Ber}(E_{\rm A}{}^M)$ and $e = \det(e_m{}^{\rm a})$.
This form of the AdS reduction rule makes clear that $\mu$-dependent terms are
the only obstruction to K\"ahler invariance. The first term yields the K\"ahler
invariant kinetic and four-fermion terms while the others provide a $\mu$-dependent
potential for the scalar fields and masses for the fermions. In components, one
finds
\begin{align}
S &= \int \rd^4x\, e\, \Bigg\{
     - \ccD_m \varphi^a g_{a \bar b}\ccD^m \bar\varphi^{\bar b}
     - \ri \chi^{\alpha a} g_{a \bar b} \nabla_{\alpha \dalpha} \bar \chi^{\dalpha \bar b}
     + \hat F^a g_{a \bar b} \bar {\hat F}^{\bar b}
     + \frac{1}{4} (\chi^a \chi^b) (\bar\chi^{\bar a} \bar\chi^{\bar b}) R_{a \bar a b \bar b}
     \eol & \quad
     - \frac{\mu}{2} (\chi^a \chi^b) \nabla_a \cK_b
     - \frac{\bar\mu}{2} (\bar\chi^{\bar a} \bar\chi^{\bar b}) \nabla_{\bar a} \cK_{\bar b}
     + \mu \hat F^a \cK_a
     + \bar \mu \bar {\hat F}^{\bar a} \cK_{\bar a}
     + 3 \mu \bar \mu \cK \Bigg\}~.
\end{align}
We have defined components in the conventional way
\begin{align}
\varphi^a := \phi^a \vert, \qquad
\chi_\alpha^b := \frac{1}{\sqrt 2} \cD_\alpha \phi^b\vert, \qquad
F^a := -\frac{1}{4} \cD^2 \phi^a\vert
\end{align}
and have made use of the quantity
\begin{align}
\hat F^a := F^a - \frac{1}{2} \Gamma^a{}_{bc} \chi^b \chi^c
\end{align}
which transforms covariantly under reparametrizations.
The component AdS derivative is given by the $\theta$-independent
piece of the superspace vector derivative,
\begin{align}
\ccD_m := e_m{}^{\rm a} \cD_{\rm a}\vert~.
\end{align}
The reparametrization-covariant derivative acts on the fermions, for example, as
\begin{align}
\nabla_{\alpha \dalpha} \bar\chi^{\dalpha \bar b} := \ccD_{\alpha \dalpha} \bar\chi^{\dalpha \bar b}
     + \Gamma^{\bar b}{}_{\bar c \bar d} \ccD_{\alpha \dalpha} \bar\varphi^{\bar c} \bar\chi^{\dalpha \bar d}~;
\end{align}
their masses are given by reparametrization-covariant field derivatives of $\cK$
\begin{align}
\nabla_a \cK_b := \partial_a \cK_b - \Gamma^c{}_{ab} \cK_c~,\qquad \cK_a := \nabla_a \cK = \partial_a \cK~.
\end{align}
This action is invariant under the $\N=2$ supersymmetry transformations
\begin{subequations}
\begin{align}
\delta \varphi^a &= \sqrt 2 \left(\lambda\chi^a + \omega^a{}_{\bar b}\, \bar\eps \bar\chi^{\bar b} \right) \\
\delta \chi_\alpha^a + \Gamma^a{}_{bc} \delta \varphi^b \chi_\alpha^c 
     &= \sqrt 2 \left(\lambda_\alpha \hat F^a - \eps_\alpha \omega^a{}_{\bar b} \bar {\hat F}^{\bar b}\right)
     + \ri \sqrt 2 \left(\bar\lambda^\dalpha \ccD_{\alpha \dalpha} \varphi^a
     - \bar\eps^\dalpha \omega^a{}_{\bar b} \ccD_{\alpha \dalpha} \bar\varphi^{\bar b}\right)\\
\delta \hat F^a + \Gamma^a{}_{bc} \delta \varphi^b \hat F^c
     &= -\bar\mu \sqrt 2 (\lambda \chi^a
     + \omega^a{}_{\bar b} \,\bar\eps \bar\chi^{\bar b})
     + \ri \sqrt 2 \left(
          \bar\lambda_\dalpha \nabla^{\dalpha \alpha} \chi_\alpha^a
          + \omega^a{}_{\bar b} \eps^\alpha \nabla_{\alpha \dalpha} \bar\chi^{\dalpha \bar b}\right)
     \eol & \quad
     + \frac{1}{\sqrt 2} \left(R_{c \bar c}{}^a{}_b \bar\lambda \bar\chi^{\bar c} \chi^c \chi^b
     - \omega^a{}_{\bar b} R_{c \bar c}{}^{\bar b}{}_{\bar d}
     \,\eps \chi^c \bar\chi^{\bar c} \bar\chi^{\bar d}\right)
\end{align}
\end{subequations}
where the spinor supersymmetry parameters $\lambda_\alpha$ and $\eps_\alpha$ obey the
AdS Killing spinor equations
\begin{subequations}
\begin{align}
& \ccD_{\a ( \ad }\bar \l_{\bd)} =0~, \qquad \ccD_{\alpha \dalpha} \bar\lambda^\dalpha = 2\ri \bar\mu \lambda_\alpha~, \\
&  \ccD_{\a ( \ad }\bar \e_{\bd)} =0~, \qquad
\ccD_{\alpha \dalpha} \bar\eps^\dalpha = 2\ri \mu \eps_\alpha~.
\end{align}
\end{subequations}
In addition, the $\rm O(2)$ rotation of AdS acts on the fields as
\begin{subequations}
\begin{align}
\delta \varphi^a &= - 2 \frac{\xi}{|\mu|} \omega^a{}_{\bar b} \bar {\hat F}^{\bar b} \\
\delta \chi_\alpha^a + \Gamma^a{}_{bc} \delta \varphi^b \chi_\alpha^c 
     &= - 2 \ri \frac{\xi}{|\mu|}\omega^a{}_{\bar b} \nabla_{\alpha \dalpha} \bar \chi^{\dalpha \bar b}
     + \frac{\xi}{|\mu|} \omega^{a b} R_{b\bar b c \bar c} \,
          \,\bar\chi^{\bar b} \bar\chi^{\bar c}  \,\chi_\alpha^c \\
\delta \hat F^a + \Gamma^a{}_{bc} \delta \varphi^b \hat F^c
     &= 6 \bar\mu \frac{\xi}{|\mu|} \omega^a{}_{\bar b} \bar {\hat F}^{\bar b}
     - 2 \frac{\xi}{|\mu|} \omega^a{}_{\bar b} \nabla_{\rm a} \ccD^{\rm a} \bar\varphi^{\bar b}
     + 2\ri \frac{\xi}{|\mu|} \omega^{a b}R_{b \bar c d \bar d}
          \chi_{\alpha}^d \bar\chi_\dalpha^{\bar d} \ccD^{\dalpha \alpha} \bar\varphi^{\bar c}
     \eol & \quad 
%      + \frac{\xi}{|\mu|} \omega^{a b} R_{b \bar c d \bar d}
%           \bar \chi^{\bar c} \bar \chi^{\bar d}\, {\hat F}^{d}
%      - \frac{\xi}{|\mu|} R_{c \bar c}{}^a{}_b \chi^b \chi^c \omega^{\bar c}{}_d \hat F^d
     + \frac{\xi}{|\mu|} R^a{}_{b \bar c c}
          \left(
          \omega^{\bar c}{}_d \chi^d \chi^c
          - \omega^c{}_{\bar d} \bar\chi^{\bar d}\bar\chi^{\bar c}
          \right) \hat F^b
     \eol & \quad 
     - \frac{1}{2} \frac{\xi}{|\mu|} \omega^{a b} \nabla_c R_{b \bar c d \bar d}
          \chi^c \chi^d \bar\chi^{\bar c} \bar\chi^{\bar d}
\end{align}
\end{subequations}
The \emph{combination} of supersymmetry and $\rm O(2)$ transformations closes off-shell.

Integrating out the auxiliary field gives $\hat F^b = -\bar\mu g^{b \bar b} \cK_{\bar b}$
and the action becomes
\begin{align}
S &=  \int \rd^4x\, e\, \Bigg\{
     - \ccD_m \varphi^a g_{a \bar b}\ccD^m \bar\varphi^{\bar b}
     - \ri \chi_\alpha^a g_{a \bar b} \nabla_{\alpha \dalpha} \bar \chi^{\dalpha \bar b}
     + \frac{1}{4} (\chi^a \chi^b) (\bar\chi^{\bar a} \bar\chi^{\bar b}) R_{a \bar a b \bar b}
     \eol & \quad
     - \frac{\mu}{2} (\chi^a \chi^b) \nabla_a \cK_b
     - \frac{\bar\mu}{2} (\bar\chi^{\bar a} \bar\chi^{\bar b}) \nabla_{\bar a} \cK_{\bar b}
     - \mu \bar \mu g^{a \bar b} \cK_a \cK_{\bar b} + 3 \mu \bar \mu \cK
     \Bigg\}~.
\end{align}
The second line can be rewritten in terms of the Killing vector
$V^a := \dfrac{\mu}{2 |\mu|} \omega^{a b} \cK_b$ as
\begin{align}
S &=  \int \rd^4x\, e\, \Bigg\{
     - \ccD_m \varphi^a g_{a \bar b}\ccD^m \bar\varphi^{\bar b}
     - \ri \chi_\alpha^a g_{a \bar b} \nabla_{\alpha \dalpha} \bar \chi^{\dalpha \bar b}
     + \frac{1}{4} (\chi^a \chi^b) (\bar\chi^{\bar a} \bar\chi^{\bar b}) R_{a \bar a b \bar b}
     \eol & \quad
     + |\mu| (\chi^a \chi^b) \omega_{bc} \nabla_a V^c
	+ |\mu| (\bar\chi^{\bar a} \bar\chi^{\bar b}) \bar\omega_{\bar b \bar c} \nabla_{\bar a} V^{\bar c}
     - 4 \mu \bar \mu g_{a \bar b} V^a V^{\bar b} + 3 \mu \bar \mu \cK
     \Bigg\}~.
\end{align}
Because $\cK$ appears explicitly in the potential, it must be a globally-defined
function (up to at most a constant shift). 

Using the equations of motion, the supersymmetry and $\rm O(2)$ transformations may
be written entirely in terms of geometric quantities,
\begin{subequations}
\begin{align}
\delta \varphi^a &= \sqrt 2 \left(\lambda \chi^a + \omega^a{}_{\bar b}
     \bar\eps \bar\chi^{\bar b} \right)
     + 4 \xi V^a \\
\delta \chi_\alpha^a + \Gamma^a{}_{bc} \delta \varphi^b \chi_\alpha^c &=
     \ri \sqrt 2 \left(\bar\lambda^\dalpha \ccD_{\alpha \dalpha} \varphi^a
     - \omega^a{}_{\bar b} \,\eps^\dalpha \ccD_{\alpha \dalpha} \bar\varphi^{\bar b}\right)
     \eol & \quad
     - 2 \sqrt 2 |\mu| \left(\lambda_\alpha \omega^a{}_{\bar b}V^{\bar b} - \eps_\alpha V^{a}\right)
     + 4\xi \,\chi_\alpha^b \nabla_b V^a~.
\end{align}
\end{subequations}

\section{Discussion}

In this paper, we have constructed the most general $\N=2$ supersymmetric
nonlinear sigma-model in AdS in terms of $\N=1$ chiral superfields.
As in the rigid
supersymmetric case, the  target space of the sigma-model must be hyperk\"ahler.
However, the AdS supersymmetry imposes some additional geometric restrictions.
The hyperk\"ahler target space $\cM$ must be such that (i) the K\"ahler two-form 
 $ \O=\ri \,g_{a \bar b} \, \rd \f^a \wedge \rd \bar \f^{\bar b}$,  which is associated with 
 the complex structure $J_3$ used in the $\cN=1$ superspace formulation, is exact 
 (and hence the target space is non-compact); (ii) $\cM$ possesses a Killing vector defined by 
(\ref{vector_field}) which rotates the three complex structures, eq. \eqref{eq_CSrotate}.
It should be pointed out that the exactness of $\O$ is a general feature 
of $\cN=1$ supersymmetric sigma-models in AdS, as demonstrated earlier 
in \cite{Adams:2011vw} and \cite{Festuccia:2011ws}.

The condition that $\cM$ must possess a certain Killing vector has in fact a simple physical
explanation. As compared with the
$\cN=2$ super-Poincar\'e group, its AdS counterpart $\rm OSp(2|4)$  includes an
additional one-parameter symmetry which is the group of O(2) rotations.  Invariance
under this symmetry proves to require the existence of a Killing vector in the target space.

A natural question to ask is whether a given hyperk\"ahler manifold with 
the properties described can be the target space of an $\N=2$ sigma-model in AdS.
Recall that if a hyperk\"ahler manifold possesses a Killing vector $V^\mu$ holomorphic
with respect to a complex structure, say $J_1$, then one can easily show that
\begin{align}
V^\mu = \frac{1}{2} J_1{}^\mu{}_\nu \nabla^\nu \cK
\end{align}
for a real \emph{Killing potential} $\cK$.\footnote{In the basis where
$J_1 = \textrm{diag}(\ri \mathbbm 1_n,-\ri \mathbbm 1_n)$,
this reduces to the usual definition of a \emph{Killing potential} \cite{BaggerWitten}
(aside from an additional numerical factor).}
However, if in addition we make the assumption
that $V^\mu$ rotates the other two complex structures, i.e.
\begin{align}
\cL_V J_1 = 0~,\qquad \cL_V J_2 = -J_3~,\qquad \cL_V J_3 = +J_1
\end{align}
then it is a simple exercise to show that
\begin{align}
g_{\mu \nu} = \frac{1}{2} (\delta_\mu{}^\rho \delta_\nu{}^\sigma + J_3{}_\mu{}^\rho J_3{}_\nu{}^\sigma) \nabla_\rho \nabla_\sigma \cK
\end{align}
or equivalently (in complex coordinates where $J_3$ is diagonalized)
\begin{align}
g_{a b} = 0~,\qquad g_{a \bar b} = \partial_a \partial_{\bar b} \cK~.
\end{align}
In other words, the function $\cK$ is not only the Killing potential
with respect to $J_1$, but \emph{also} the \emph{K\"ahler potential}
with respect to $J_3$. In fact, it is the K\"ahler potential
with respect to \emph{any} complex structure orthogonal to $J_1$.

We are thus led to the following simple prescription for generating an
$\N=2$ nonlinear sigma-model in AdS from a given hyperk\"ahler manifold.
If the hyperk\"ahler manifold admits some Killing vector $V^\mu$ which
rotates the complex structures (necessarily leaving one of
them invariant) then one constructs a Killing potential $\cK$ with respect to the
invariant complex structure. The resulting function is the \emph{K\"ahler potential}
and, indeed, the superfield Lagrangian when written in the basis where one
of the orthogonal complex structures is diagonalized. For this prescription
to be consistent, the K\"ahler form associated with the diagonalized 
complex structure must be exact (hence the hyperk\"ahler manifold must be
non-compact) and the function $\cK$ must be globally defined.

It is quite intriguing that some of the properties we have discussed
have recently been independently discovered in the context of supersymmetric
nonlinear sigma-models in ${\rm AdS}_5$ \cite{BX2011, BaggerLi}.\footnote{The paper \cite{BX2011}, which
appeared on the preprint arXiv shortly after the first version of this
paper, was the first to note that the Killing vector in AdS$_5$ was holomorphic
while \cite{BaggerLi} later noted that it rotated the complex structures.
Subsequently we discovered the same features in AdS$_4$, where they are
more hidden.} In particular, a Killing vector $V^\mu$ appears
which rotates the complex structures while leaving one of them invariant.
However, in these models, it is the invariant complex structure which is
diagonalized, and so $V^\mu$ is holomorphic in the usual sense. This undoubtedly is related
to the fact that in these models the five-dimensional space is foliated with
flat four-dimensional subspaces. More precisely, the Killing vector turns out to be
holomorphic due to a certain imbedding of the hypermultiplets into 4D $\N=1$ chiral superfields.

The remarkable feature of our construction is that the $\N=2$ supersymmetry
algebra closes off the mass shell, for the most general  $\N=2$ supersymmetric
nonlinear sigma-model in AdS realized in terms of  $\N=1$ chiral superfields. 
This is a new type of structure that has no analogue in Minkowski space. 
Indeed, in order to have off-shell supersymmetry for general $\cN=2$ nonlinear sigma-models 
in Minkowski space, one has to use the harmonic \cite{GIKOS,GIOS}
or the projective \cite{KLR,LR-projective} superspace approaches
in which the off-shell hypermultiplet realizations involve  an infinite number of auxiliary fields.
In our construction, the hypermultiplet is described using a minimal realization of two ordinary $\cN=1$ chiral superfields with $8+8$ degrees of freedom. One may wonder why the structure of supersymmetry 
transformations in AdS differs so drastically from that in Minkowski space. The origin of this difference can be traced back to the explicit form of the AdS superfield parameter (\ref{SUSY_parameter}). One can see that 
the leading component of $\ve$ is not analytic in the cosmological constant $|\m|$, 
which is similar to the well-known non-analyticity of the cubic interaction of massless higher spin fields 
in AdS \cite{FV}. Thus the parameter $\ve $ does not admit a smooth limit to Minkowski space.
On the other hand, from the work of \cite{HKLR,BX,K-duality} it is  known that in the case
of $\cN=2$ nonlinear sigma-models in Minkowski space 
one has to deal with a superfield parameter of the form
\bea
\ve = \t + \e^\a \q_\a + \bar \e_\ad \bar \q^\ad  ~, \qquad \t=\text{const}~, \qquad
\e^\a =\text{const}~.
\eea
Here the bosonic parameter $\t$ generates a central charge transformation which can be shown to be a trivial 
symmetry (i.e. it coincides with the identity transformation on-shell). 
This transmutation of the physical O(2) symmetry, which is  generated by the parameter $\x$ 
in  (\ref{SUSY_parameter}), 
into a trivial $\t$-symmetry is another manifestation of non-analyticity in the cosmological constant.

Off-shell supersymmetry is also characteristic of the gauge models for massless higher spin 
$\cN=2$ supermultiplets in AdS constructed in \cite{GKS} using $\cN=1$ superfields. 
Since those theories are linearized, one may argue that their  off-shell supersymmetry is not really impressive.
However, now we have demonstrated that the formulation of the most general nonlinear $\cN=2$ supersymmetric 
sigma-models in terms of $\cN=1$ chiral superfields is also off-shell. 
This gives us some evidence to believe that, say, general $\cN=2$ super Yang-Mills theories in AdS 
possess an off-shell formulation in which the hypermultiplet is realized in terms of two chiral superfields.
If this conjecture is correct, there may be nontrivial implications for quantum effective actions.

The off-shell structure of our $\cN=2$ nonlinear sigma-models in AdS 
implies that there should exist a \emph{manifestly} $\N=2$ supersymmetric 
formulation in AdS with the same finite set of auxiliary fields we have
found. It would be of interest to develop such a formulation.\\

\noindent
{\bf Acknowledgements:}\\
This work is supported in part by the Australian Research Council 
and by a UWA Research Development Award. D.B. would like to thank
Jon Bagger for correspondence.
\\

\noindent
{\bf Note added in proof:}\\
After this article was accepted for publication, the authors learned that
hyperk\"ahler geometries possessing Killing vectors which rotate the complex
structures were considered previously by Hitchin et al. \cite{HitchinKLR}.
\\

\noindent

\footnotesize{

}

\end{document}